\documentclass[ 12pt, onecolumn, journal]{IEEEtran}

\usepackage{booktabs}
\usepackage{multirow} 
\usepackage{cite}
\usepackage[cmex10]{amsmath}
\usepackage{amssymb} 
\usepackage{amsfonts}
\usepackage{amsthm}
\usepackage{bm} 
\usepackage{graphicx,epsfig} 
\usepackage{setspace}
\usepackage{algorithm}
\usepackage{algorithmicx}
\usepackage{algpseudocode}
\newtheorem{lem}{Lemma}

\newtheorem{Col}{Corollary}
\newtheorem{Assump}{Assumption}
\usepackage[margin=1 in]{geometry}

\newcommand {\eqdef}{\ensuremath{\stackrel{{\scriptscriptstyle\triangle}}{=}}}  

\begin{document}

\title{MIMO Interference Alignment Over Correlated Channels with Imperfect CSI}
\author{Behrang~Nosrat-Makouei,~\IEEEmembership{Student Member,~IEEE,}
        Jeffrey~G.~Andrews,~\IEEEmembership{Senior Member,~IEEE,}
        and Robert~W.~Heath,~Jr.,~\IEEEmembership{Senior Member,~IEEE}
\thanks{The authors are with the Department of Electrical and Computer Engineering, The University of Texas at Austin, Austin, TX 78712 USA (e-mail:
behrang.n.m@mail.utexas.edu; jandrews@ece.utexas.edu; rheath@ece.utexas.edu).}
\thanks{This work was presented in part at IEEE International Symposium on Information Theory, 2010 \cite{BehrangISIT2010}.}
\thanks{October 12, 2010}}
\maketitle

\begin{abstract}
Interference alignment (IA), given uncorrelated channel components and perfect channel state information, obtains the maximum degrees of freedom in an interference channel. Little is known, however, about how the sum rate of IA behaves at finite transmit power, with imperfect channel state information, or antenna correlation. This paper provides an approximate closed-form signal-to-interference-plus-noise-ratio (SINR) expression for IA over multiple-input-multiple-output (MIMO) channels with imperfect channel state information and transmit antenna correlation.
Assuming linear processing at the transmitters and zero-forcing receivers, random matrix theory tools are utilized to derive an approximation for the post-processing SINR distribution of each stream for each user. Perfect channel knowledge and i.i.d. channel coefficients constitute special
cases. This SINR distribution not only allows easy calculation of useful
performance metrics like sum rate and symbol error rate, but also permits a realistic comparison of IA with other transmission techniques. More specifically, IA is compared with spatial multiplexing and beamforming and it is shown that IA may not be optimal for some performance criteria.
\end{abstract}
\section{Introduction}
\indent Many important wireless communication scenarios can be modeled using an interference channel. Examples include interfering base stations in a cellular network, wireless local area networks, and simultaneous transmission in mobile ad-hoc networks. The general capacity of the interference channel and the design of practical schemes approaching the known upper bounds on sum rates have been of great interest over the last 30 years. The earliest attempts to characterize the capacity region of the interference channel, inspired by the framework established by Shannon in \cite{shannon1961two}, were focused on two-user interference channels. Although the special cases of strong and very strong interference have been solved \cite{Sato1981, carleial1975case}, the general capacity of the interference channel is still an open problem. Recently, a series of attempts have been made to describe an approximation of the asymptotic sum capacity behavior known as the maximum achievable multiplexing gain or \textit{degrees of freedom} (DoF) \cite{Cadambe2008a} where focus is on the high SNR regime and the interference/broadcast characteristics of the wireless network. The DoF studies paved the way for a novel method of dealing with interference, known as \textit{{interference alignment}} (IA) \cite{Maddah-Ali2008,Cadambe2008a}.

\subsection{Recent Work and Motivation}
\indent IA uses beamforming matrices at the transmitters to align the interference to a received signal subspace such that an interference free subspace becomes available for direct signal transmission. Thus in IA, the primary goal of the transmitting nodes is to reduce their decremental effect on the unintended receivers. In contrast to other techniques of interference management such as {orthogonal access}, {decoding the interference} \cite{Han1981} or {treating the interference as noise} \cite{Etkin2008}, IA achieves the maximum DoF in a $K$-user interference channel \cite{Cadambe2008a}. 
IA has also been used successfully to characterize the maximum DoF of other network scenarios including the MIMO interference channel with time varying/frequency selective channels \cite{Cadambe2008a} or constant channel coefficients \cite{Peters2009, Tresch2009a}, the X channel \cite{Huang2009a}, the MIMO X channel \cite{Maddah-Ali2008}, X \cite{Cadambe2009a}, and MIMO X networks \cite{Jafar2008}. 

\indent Given the potential of IA, recent work has further explored its applications and limitations. Iterative and distributed algorithms for IA over constant channel coefficients are presented in \cite{Gomadam2008, Peters2009}.
The feasibility of IA over spatial dimensions is studied in \cite{Tresch2009a,Yetis2009}. Reducing the overhead associated with IA is considered in \cite{Peters2010}. 
Adaptation of IA to multi-cell networks is considered in \cite{Tresch2009,tresch2009clustered}. A method of opportunistic access in cognitive radios inspired from IA is proposed in \cite{Perlaza2008}. Using IA for secure communications is studied in \cite{Koyluoglu2008}. And finally, extending IA to relay-aided networks is considered in \cite{Mohajer2008,Khandani2009,ElAyach2009}. 

\indent On the one hand, despite the large number of practical wireless networks for which IA is being considered, IA in its original form \cite{Cadambe2008a} is only shown to be optimum for asymptotically high SNR given perfect channel state information (CSI) and i.i.d. channel coefficient values among the users and between the antennas. On the other hand, in practical systems \cite{andrews2007fundamentals}, the communicating nodes only have access to an imperfect estimate of the channel coefficients, working at intermediate SNR values is inevitable and correlation either between the nodes or especially between the antennas exists. Previous work to quantify the effect of imperfect CSI and intermediate SNR values on IA are few, and either confined to calculating bounds on achievable sum rates \cite{Tresch2009} or rely on experimental results \cite{ElAyach2009} without providing accurate quantification of key performance measures such as symbol error rate or achievable sum rate. The reason is the complicated expressions of the linear beamforming/combining filters.

\indent Our motivation for considering ZF receivers is two fold. First, although zero-forcing (ZF) receivers are asymptotically sufficient for achieving the DoF promised by IA \cite{Cadambe2008a} and they provide a simple and effective method for multiple stream detection, in the context of IA, little is known about the performance of such receivers in intermediate SNR regimes with imperfect CSI and correlated antennas/channels. Second, some other classes of receiver filters, such as minimum mean square error or regularized ZF filters, converge to the ZF receiver at high SNR which raises further question about ZF receivers at lower SNR. 

\indent Note that interference can be aligned on both the signal level space \cite{Sridharan2008} (multi-level and lattice coding) and in the signal vector space \cite{Cadambe2008a, Gou2008a,Jafar2008} (designing the beamforming and combining filters using the time/frequency/spatial dimensions). 
Furthermore, although next generation MIMO based networks \cite{andrews2007fundamentals} present the opportunity to code over frequency, time and space dimensions, lack of sensitive frequency synchronization between the nodes, potential high correlation between the channel coefficient values (IA highly relies on independence between the used channel coefficients to achieve the full multiplexing gain \cite{Cadambe2008a}) and delay-sensitive communications justifies using IA only over the spatial dimensions, i.e. IA over constant MIMO channels \cite{Tresch2009a, Peters2009}.

\subsection{Contributions}
\indent In this paper, we quantify the IA performance under imperfect CSI for constant MIMO channels with transmit correlation with ZF filters. First, using random matrix theory tools, we show that given perfect CSI at all the nodes and uncorrelated i.i.d. channel coefficients, the received SNR per stream for each user after a ZF equalizer is exponentially distributed. In other words, using a ZF receiver results in parallel single-input-single-output Rayleigh channels. Next, we show that if there exists an arbitrary Kronecker-modeled transmit correlation, using the asymptotic behavior of the eigenvectors of Wishart matrices, we can quantify its effect approximately on the received SNR distribution. Our analysis shows that the accuracy of this approximation depends on the number of antennas at each node, the transmit antenna correlation, and the transmit power. 

\indent After quantifying the effect of transmit correlation on the SNR distribution, we analyze the effect of imperfect CSI (Gauss-Markov model) on the received SNR. We show that the imperfect channel knowledge reduces the mean received SNR in proportion to the number of streams. In other words, increasing the multiplexing gain, either by using more antennas at each node or by increasing the number of transmit-receive pairs, increases the detrimental impact of imperfect channel knowledge. Moreover, we show that if the imperfection in CSI does not vanish at asymptotically high transmit powers, one can not achieve the full multiplexing gain promised by IA.
Finally we show that using the derived per-stream SNR distributions, it is possible to accurately compare an IA network with other interference management methodologies such as an orthogonal access network which is utilizing spatial multiplexing (SM) or beamforming. Our recent work on this subject \cite{BehrangISIT2010} did not include channel correlation and was restricted to quantifying the impact of imperfect CSI on the distribution of the post-processing SINR for IA over constant MIMO channels. 

\subsection{Organization and Notation}
\indent \textit{Organization}: In Section \ref{sec:PCSI} we present the system model and analyze the performance of IA given perfect channel knowledge and i.i.d. channel coefficient values. In Sections \ref{sec:Tcorr} and \ref{sec:IPCSI} we progressively introduce transmit correlation and imperfect CSI into the system model and then quantify their effects on the distribution of the  post-processing SINR of each received stream. Section \ref{sec:P2P} analyzes a point-to-point MIMO system relative to our IA configuration. Numerical experiments are presented in Section \ref{Sec:Numerical} and concluding remarks are given in Section \ref{sec:Conclusion}.

\indent \textit{Notation}: Capital and small bold letters stand for matrices and vectors. $\mathbf{A}^*$, $\mathbf{A}^T$ and $\mathbf{A}^{-1}$ are conjugate transpose, transpose and inverse of $\mathbf{A}$ respectively. $\mathbf{A}(n,m)$ is the element on the $n^{th}$ row and $m^{th}$ column of $\mathbf{A}$. $\mathbf{A}(:,m)$ is the vector of $m^{th}$ column  of $\mathbf{A}$. $\rm{tr}\big(\mathbf{A}\big)$ and $\rm{rank}(\mathbf{A})$ are trace and rank of $\mathbf{A}$, respectively. $[\mathbf{A},\mathbf{B}]$ is the matrix constructed by horizontal concatenation of matrices $\mathbf{A}$ and $\mathbf{B}$. $\rm{vec}\left(\mathbf{A}\right)$ operator stacks the columns of $\mathbf{A}$. $\lceil a \rceil$ is the smallest integer larger than or equal to $a$, $\mathbf{I}_N$ is the $N\times N$ identity matrix and $\mathbf{0}_{a\times b}$ is an $a\times b$ matrix of all zeros.

\section{IA Over a Constant MIMO Channel} \label{sec:PCSI}
\begin{figure}[t]
 \centering
\includegraphics[bb= 0 0 285 300,width=3.5in]{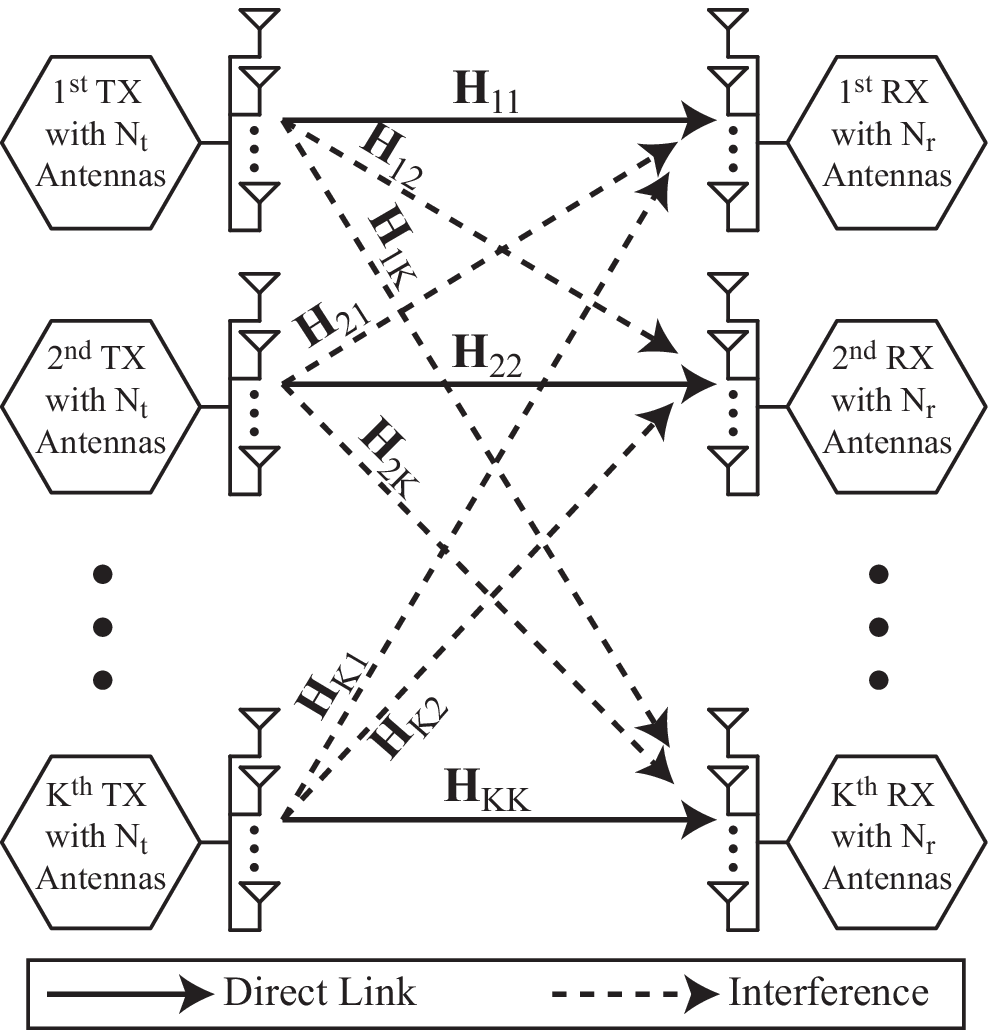}
\caption{A $K$-user MIMO interference channel where transmitters and receivers have  $N_t$ and $N_r$ antennas respectively. }
\label{fig:SystemModel}
\end{figure}

Consider the $K$-user $N_r\times N_t$ MIMO interference channel shown in Fig. \ref{fig:SystemModel}.
Transmitter $i$ encodes $d_i$ streams using a precoding matrix $\mathbf{F}_i$ which is then decoded by the $i^{th}$ receiver after processing the received signal with a combining matrix $\mathbf{W}_i$.   Assuming perfect synchronization between the nodes, the received signal at receiver $i$ can be written as
\begin{align}
\mathbf{y}_i &= \sum_{k= 1}^K \mathbf{H}_{ik}\mathbf{F}_{k}\mathbf{x}_{k} + \mathbf{z}_i,  \label{eq:RxSignal_origin}
\end{align}
where $\mathbf{z}_i$ is AWGN with elements distributed as $\mathcal{CN}(0,\sigma^2)$, transmitted streams $\mathbf{x}_i$ obey the total power constraint $\mathbb{E}\left\{\mathbf{x}_i^*\mathbf{x}_i\right\}= P$, $\mathbf{H}_{ik}$  is the matrix of the channel coefficients between receiver $i$ and transmitter $k$. Note that in constant MIMO channels, coding over frequency/time does not change the achievable DoF \cite{Cadambe2008a} and without loss of generality we can assume the channels are narrowband. 

\indent The goal of IA for the MIMO constant channel is to design the precoding and combining matrices such that the following conditions are satisfied 
\begin{align}
\left\{ \begin{array}{l}
\rm{rank}\big(\mathbf{W}_i\mathbf{H}_{ii}\mathbf{F}_i\big) = d_i \\
\mathbf{W}_i\mathbf{H}_{ik}\mathbf{F}_{k} = \mathbf{0}\ \forall k\neq i
\end{array} \right. \forall i\in\{1,\ldots,K\}. \label{eq:IAconds}
\end{align}
Except the $3$-user \cite{Cadambe2008a} and the $(N+1)$-user $N\times N$ \cite{Tresch2009a} constant MIMO channels, analytical expressions directly solving (\ref{eq:IAconds}) for the unknowns are still under investigation. There exist, however, numerical methods for both designing the precoding/combining matrices and calculating the achievable DoF \cite{Peters2009,Gomadam2008}. We focus in this paper on the alternating minimization method in\cite{Peters2009}; we leave extending the developed theory and the results to other IA precoding/combining matrix design techniques  for future work. Note that a system of IA can have multiple solutions\cite{Yetis2009} and it is possible to select a solution with specific properties for each channel instance\cite[Section VI]{ElAyach2009a}. Our analysis, however, is based on selecting a random IA solution for a given channel instance and studying the behavior of solutions with specific properties (through extreme statistics) is out of the scope of this work. 

\indent For the $i^{th}$ transmitter/receiver pair, the alternating minimization method results in a unitary transmitter beamforming matrix, $\mathbf{F}_i$, and a set of non-unique orthonormal basis for the interference subspace, i.e. columns of $\mathbf{C}_i$, such that
\begin{align}
\mathbf{F}_i^*\mathbf{F}_i = \mathbf{C}_i^*\mathbf{C}_i = \mathbf{I}\quad \forall i \in \{1,\ldots,K\}. \label{eq:PrecodeCombineConditions}
\end{align}
Note that in the alternating minimization method, the direct channel links $\mathbf{H}_{ii}$ do not appear in the computation of precoding matrices (and interference subspaces)\cite{Peters2009} and therefore the elements of $\mathbf{H}_{ii}$ are independent of $\mathbf{F}_i$ and $\mathbf{C}_i$ for $i\in\{1,\ldots,K\}$. 

\indent The next step is to design the receiver equalizers. Assuming the alternating minimization method has converged to an IA solution (see \cite{Peters2009} and \cite[Section 6]{Negro2009} for discussions on its convergence), by knowing a basis for the interference subspace at each receiver $i$, i.e. columns of $\mathbf{C}_i$, (\ref{eq:RxSignal_origin}) can be written as
\begin{align}
\mathbf{y}_i &=  \mathbf{H}_{ii}\mathbf{F}_i\mathbf{x}_i + \mathbf{C}_i\sum_{k\neq i}^K\mathbf{A}_{ik}\mathbf{x}_{k} + \mathbf{z}_i 
= \tilde{\mathbf{H}}_i\left[\begin{array}{cc}
\mathbf{x}_i \\
\sum_{k\neq i}^K\mathbf{A}_{ik}\mathbf{x}_{k} \end{array}\right] + \mathbf{z}_i, \label{eq:yiHeff}
\end{align}
where $\mathbf{A}_{ik}$ determines the interference from transmitter $k$ at receiver $i$ and is given by $\mathbf{C}_i\mathbf{A}_{ik} = \mathbf{H}_{ik}\mathbf{F}_{k}$ and $\tilde{\mathbf{H}}_i  \eqdef \big[\mathbf{H}_{ii}\mathbf{F}_i,\mathbf{C}_i\big]$ is the effective channel at receiver $i$. Note that any other orthonormal basis for the interference subspace is related to $\mathbf{C}_i$ through a unitary mapping and, by appropriately transforming $\mathbf{A}_{ik}$, we can restrict our attention to $\mathbf{C}_i$ without affecting the forthcoming discussions.
Moreover, as $\mathbf{H}_{ii}$ is independent of $\mathbf{C}_i$ and  $\mathbf{F}_i$ (through construction), the columns of the $N_r\times d_i$ matrix of $\mathbf{H}_{ii}\mathbf{F}_i$ do not completely lie in the subspace spanned by columns of the $N_r\times(N_r-d_i)$ matrix of $\mathbf{C}_i$, i.e. the $N_r\times N_r$ matrix of $\big[\mathbf{H}_{ii}\mathbf{F}_i,\mathbf{C}_i\big]$ is full rank. In addition, given that IA is feasible and that transmitters are communicating the maximum number of allowed streams, $\mathbf{F}_i$ and $\mathbf{C}_i$ are always of dimension $N_t\times d_i$ and $N_r\times(N_r-d_i)$ and $\tilde{\mathbf{H}}_i$ is always square.
Therefore, the ZF equalizer at receiver $i$ will be
\begin{align}
\mathbf{W}_i^{ZF} = \big[\mathbf{I}_{d_i},\mathbf{0}_{d_i\times (N_r-d_i)}\big]\tilde{\mathbf{H}}_i^{-1}. \label{eq:ZFpcsi}
\end{align}
Define $\mathbf{B}_i = \big[\mathbf{I}_{d_i},\mathbf{0}_{d_i\times (N_r-d_i)}\big]$. Applying the ZF receiver given by (\ref{eq:ZFpcsi}) to the received signal given in (\ref{eq:yiHeff}) leads to an expression for the SNR of the $n^{th}$ stream at receiver $i$ 
\begin{align}
\gamma_{i,n} = \frac{\gamma_o/d_i}
{\left[\mathbf{B}_i(\tilde{\mathbf{H}}_i^*\tilde{\mathbf{H}}_i)^{-1}\mathbf{B}_i^*\right]_{n,n}} \ , \label{eq:SNR_NoError_temp}
\end{align}
where $\gamma_o = \frac{P}{\sigma^2}$. Using Lemma \ref{Lem:PartitionedInverse}, the denominator of (\ref{eq:SNR_NoError_temp}) can be simplified into an expression better suited for statistical analysis. 

\begin{lem}\label{Lem:PartitionedInverse}  
Assuming IA is feasible, the denominator of (\ref{eq:SNR_NoError_temp}) simplifies to $\left[\mathbf{F}_i^*\mathbf{H}_{ii}^*\left(\mathbf{I}_{N_r} - \mathbf{C}_i\mathbf{C}_i^*\right)\mathbf{H}_{ii}\mathbf{F}_i\right]_{n,n}$.
\end{lem}
\begin{IEEEproof}
Given that $\tilde{\mathbf{H}}_i = \big[\mathbf{H}_{ii}\mathbf{F}_i,\mathbf{C}_i\big]$
\begin{align}
\tilde{\mathbf{H}}_i^*\tilde{\mathbf{H}}_i=\big[\mathbf{H}_{ii}\mathbf{F}_i,\mathbf{C}_i\big]^*\big[\mathbf{H}_{ii}\mathbf{F}_i,\mathbf{C}_i\big]  
= \left[\begin{array}{cc}
\mathbf{\Gamma}_1& \mathbf{\Gamma}_2 \\ \mathbf{\Gamma}_3 & \mathbf{\Gamma}_4 \end{array} \right] \ , \label{eq:TempInSimplifyingDenomProof}
\end{align}
where ${\mathbf{\Gamma}_1 = \mathbf{F}_i^*\mathbf{H}_{ii}^*\mathbf{H}_{ii}\mathbf{F}_i}$,  ${\mathbf{\Gamma}_2 = \mathbf{F}_i^*\mathbf{H}_{ii}^*\mathbf{C}_i}$,  ${\mathbf{\Gamma}_3 = \mathbf{C}_i^*\mathbf{H}_{ii}\mathbf{F}_i}$ and ${\mathbf{\Gamma}_4 = \mathbf{C}_i^*\mathbf{C}_i = \mathbf{I}_{N_r-d_i}}$. Noting that left and right multiplication of any matrix by $\mathbf{B}_i$ and $\mathbf{B}_i^*$ keeps the first $d_i$ rows and the first $d_i$ columns of that matrix, if $\left(\mathbf{\Gamma}_1 - \mathbf{\Gamma}_2(\mathbf{\Gamma}_4)^{-1}\mathbf{\Gamma}_3\right)$ is not singular, the inverse of (\ref{eq:TempInSimplifyingDenomProof}) simplifies to 
$\left(\mathbf{\Gamma}_1 - \mathbf{\Gamma}_2(\mathbf{\Gamma}_4)^{-1}\mathbf{\Gamma}_3\right)^{-1}$ \cite[Section 3.5.3]{Lutkepohl1996}. As $ \left(\mathbf{I}_{N_r} - \mathbf{C}_i\mathbf{C}_i^*\right)$ is a projection matrix and hence an (Hermitian) idempotent matrix
\begin{align}
\mathbf{\Gamma}_1 - \mathbf{\Gamma}_2(\mathbf{\Gamma}_4)^{-1}\mathbf{\Gamma}_3 &= \mathbf{F}_i^*\mathbf{H}_{ii}^*\mathbf{H}_{ii}\mathbf{F}_i -\mathbf{F}_i^*\mathbf{H}_{ii}^*\mathbf{C}_i\mathbf{C}_i^*\mathbf{H}_{ii}\mathbf{F}_i \nonumber \\
& = \mathbf{F}_i^*\mathbf{H}_{ii}^*\left(\mathbf{I}_{N_r} - \mathbf{C}_i\mathbf{C}_i^*\right)\mathbf{H}_{ii}\mathbf{F}_i = 
\mathbf{F}_i^*\mathbf{H}_{ii}^*\tilde{\mathbf{C}}_i\mathbf{H}_{ii}\mathbf{F}_i 
 = \mathbf{F}_i^*\mathbf{H}_{ii}^*\tilde{\mathbf{C}}_i\tilde{\mathbf{C}}_i\mathbf{H}_{ii}\mathbf{F}_i , \label{eq:NonSingularCond}
\end{align}
where $\tilde{\mathbf{C}}_i = \left(\mathbf{I}_{N_r} - \mathbf{C}_i\mathbf{C}_i^*\right)$. Furthermore, $\tilde{\mathbf{C}}_i$ is a projection matrix into the $d_i$ dimensional null-space of the interference subspace at the $i^{th}$ receiver constituting a combining matrix which satisfies (\ref{eq:IAconds}). Therefore, ${\rm{rank}\left( \tilde{\mathbf{C}}_i\mathbf{H}_{ii}\mathbf{F}_i\right) = d_i }$ and the $d_i\times d_i$ matrix in (\ref{eq:NonSingularCond}) is nosingular.
\end{IEEEproof}
Using Lemma \ref{Lem:PartitionedInverse}, (\ref{eq:SNR_NoError_temp}) simplifies to
\begin{align}
\gamma_{i,n} &= \frac{\gamma_o/d_i}
{\left[\left(\mathbf{F}_i^*\mathbf{H}_{ii}^*\tilde{\mathbf{C}}_i\mathbf{H}_{ii}\mathbf{F}_i\right)^{-1}\right]_{n,n}}.  \label{eq:SNR_NoError}
\end{align}
Define $\mathbf{\Delta}_i = \mathbf{F}_i^*\mathbf{H}_{ii}^*$. Lemma \ref{Lem:WishDist_Pre} gives the distribution of $\mathbf{\Delta}_i\mathbf{\Delta}_i^*$ which is then used in Lemma \ref{Lem:WishDist} to find the distribution of $\mathbf{\Delta}_i\tilde{\mathbf{C}_i}\mathbf{\Delta}_i^*$.
\begin{lem}\label{Lem:WishDist_Pre}  
If the channel coefficients are i.i.d. zero-mean unit-variance Gaussian random variables, $\mathbf{H}_{ii}$ is independent of $\mathbf{F}_i$, (\ref{eq:PrecodeCombineConditions}) holds, then the $d_i\times d_i$ random matrix $\mathbf{\Delta}_i\mathbf{\Delta}_i^*$ is a complex Wishart matrix with $N_r$ degrees of freedom and covariance matrix $\mathbf{I}_{d_i}$.
\end{lem}
\begin{IEEEproof}In the alternating minimization algorithm, the columns of $\mathbf{F}_i$ are $d_i$ eigenvectors (corresponding to the $d_i$ least significant eigenvalues) of the $N_t\times N_t$ matrix $\sum_{k\neq i}^{K}\mathbf{H}_{ki}^*\left(\mathbf{I}-\mathbf{C}_k\mathbf{C}_k^*\right)\mathbf{H}_{ki}$.
Let the eigenvalue decomposition of  $\sum_{k\neq i}^{K}\mathbf{H}_{ki}^*\left(\mathbf{I}-\mathbf{C}_k\mathbf{C}_k^*\right)\mathbf{H}_{ki}$ be $\mathbf{U}_i\mathbf{\Sigma}_i\mathbf{U}_i^*$ where $\mathbf{\Sigma}_i$ is a diagonal matrix holding the eigenvalues sorted in ascending order and $\mathbf{U}_i$ is a unitary matrix holding the corresponding eigenvectors. 
Then $
\mathbf{F}_i = \mathbf{U}_i\big[\mathbf{I}_{d_i},\mathbf{0}_{d_i\times (N_t-d_i)}\big]^T$. Therefore $\mathbf{\Delta}_i = \big[\mathbf{I}_{d_i},\mathbf{0}_{d_i\times (N_t-d_i)}\big]\mathbf{U}_i^*\mathbf{H}_{ii}^*$ and following the unitarily invariance property of $\mathbf{H}_{ii}^*$, $\big[\mathbf{I}_{d_i},\mathbf{0}_{d_i\times (N_t-d_i)}\big]\mathbf{U}_i^*\mathbf{H}_{ii}^*$ has $N_r$ columns of size $d_i$ with each element distributed as $\mathcal{CN}(0,1)$. It follows that the $d_i\times d_i$ matrix of $\mathbf{\Delta}_i\mathbf{\Delta}_i^*$ has a central Wishart distribution with $N_r$ degrees of freedom and covariance matrix $\mathbf{I}_{d_i}$. 
\end{IEEEproof}

Using Lemma \ref{Lem:WishDist_Pre} and noting that $\tilde{\mathbf{C}}_i$ is a projection matrix into a sub-space of dimension $d_i$, Lemma \ref{Lem:WishDist} gives the distribution of $\mathbf{\Delta}_i\tilde{\mathbf{C}}_i\mathbf{\Delta}_i^*$.

\begin{lem}\label{Lem:WishDist} 
If the $d\times d$ matrix $\mathbf{\Delta}\mathbf{\Delta}^*$ has a central Wishart distribution with $N$ degrees of freedom and covariance matrix $\mathbf{I}_{d}$ and $\tilde{\mathbf{C}}$ is a projection matrix into a sub-space of dimension $c\leq N$, the $d\times d$ matrix of $ \mathbf{\Delta}\tilde{\mathbf{C}}\mathbf{\Delta}^*$ has a central Wishart distribution with $c$ degrees of freedom and covariance matrix $\mathbf{I}_{d}$.
\end{lem}
\begin{IEEEproof}
As a projection matrix into a sub-space, $\tilde{\mathbf{C}}$ can be rewritten as $\mathbf{U}\mathbf{\Sigma}\mathbf{U}^*$, where $\mathbf{U}$ is a unitary matrix and $\mathbf{\Sigma}$ holds the $c$ unity eigenvalues of $\tilde{\mathbf{C}}$ on its main diagonal and zeros elsewhere. Proof follows from  unitarily invariance of columns of $\mathbf{\Delta}$ and idempotent property of $\mathbf{\Sigma}$ \cite[Theorem 3.4.4]{Mardia1979}.
\end{IEEEproof}

\noindent From Lemma \ref{Lem:WishDist} it follows that 
the SNR of each stream given by (\ref{eq:SNR_NoError}) is exponentially distributed\cite{gore2002performance}, i.e. 
\begin{align}
f(\gamma_{i,n})=\frac{d_i}{\gamma_o}
\exp\left(-\frac{d_i\gamma_{i,n}}{\gamma_o}\right). \label{eq:pdfNoError}
\end{align}
By using a ZF receiver, the system effectively consists of $\displaystyle{\sum_{m=1}^Kd_m}$ parallel Rayleigh point-to-point links.

\section{Transmit Antenna Correlation} \label{sec:Tcorr}
A more general channel model includes spatial correlation between the channel elements. In this section we suppose that there is transmit spatial correlation, thus 
\begin{align}
\mathbf{H}_{ik} = \mathbf{H}_{ik}^w\mathbf{R}_t^{1/2}\quad \forall i,k\in\{1,\ldots,K\}, \label{eq:AntennaCorrModel}
\end{align}
where $\mathbf{H}^w\sim \mathcal{CN}(\mathbf{0},\mathbf{I})$ and $\mathbf{R}_t$ is a constant Hermitian positive semidefinite (PSD) matrix \cite{Shiu2000}.
Note that when the receive spatial correlation, $\mathbf{R}_r$, is not an identity matrix, neither the columns nor the rows of $\mathbf{H}_{ii}$ are independent and analyzing the distribution of $\mathbf{H}_{ii}^*\tilde{\mathbf{C}}_i\mathbf{H}_{ii}$ (which appears in the denominator of SNR expressions) is more challenging. Therefore, including receive correlation is a topic of future work. For the initial analysis, we assume that the spatial correlation is the same for all channels in the interference network; we relax this assumption at the end of the section. 

\indent Assume IA is done over the instantaneous channels as before. Note that the feasibility of a system of IA is a function of number of equations and variables in (\ref{eq:IAconds})\cite{Yetis2009}. Therefore, as long as $\mathbf{R}_t$ is not rank deficient, transmit correlation will not change the achievability of IA. Replacing (\ref{eq:AntennaCorrModel}) in (\ref{eq:SNR_NoError}) gives
\begin{align}
\gamma_{i,n}^c &= \frac{\gamma_o/d_i}
{\left[\left(\mathbf{F}_i^*(\mathbf{R}_t^{1/2})^*(\mathbf{H}^w_{ii})^*
\tilde{\mathbf{C}}_i\mathbf{H}^w_{ii}\mathbf{R}_t^{1/2}\mathbf{F}_i\right)^{-1}\right]_{n,n}}  = \frac{\gamma_o/d_i}
{ \left[\left(\mathbf{\tilde{\Delta}}_i\tilde{\mathbf{C}}_i
\mathbf{\tilde{\Delta}}_i^*\right)^{-1}\right]_{n,n}} \ ,
\label{eq:SNR_NoErrorExpanded_AntennaCorr}
\end{align}
where $\mathbf{\tilde{\Delta}}_i = \mathbf{F}_i^*(\mathbf{R}_t^{1/2})^*(\mathbf{H}^w_{ii})^*$. $\mathbf{\tilde{\Delta}}_i$ has i.i.d. columns with covariance matrix $\tilde{\mathbf{R}}_i$ defined as
\begin{align}
\tilde{\mathbf{R}}_i=\rm{cov}(\mathbf{\tilde{\Delta}}_i(:,n),\mathbf{\tilde{\Delta}}_i(:,n)) &= 
\mathbb{E}\left\{ \mathbf{F}_i^*(\mathbf{R}_t^{1/2})^*(\mathbf{H}^w_{ii}(:,n))^*\mathbf{H}^w_{ii}(:,n)
\mathbf{R}_t^{1/2}\mathbf{F}_i\right\}  
=\mathbb{E}\left\{\mathbf{F}_i^*\mathbf{R}_t\mathbf{F}_i\right\}. \label{eq:CovOfDelta}
\end{align}
Assuming $\tilde{\mathbf{R}}_i$ is known, similar to (\ref{eq:pdfNoError}), the SINR of the $n^{th}$ stream at receiver $i$ given by (\ref{eq:SNR_NoErrorExpanded_AntennaCorr}) is exponentially distributed\cite{gore2002performance} as
\begin{align}
f(\gamma_{i,n}^c)=\frac{d_i\sigma^2_{i,n}}{\gamma_o}
\exp\left(-\frac{d_i\sigma^2_{i,n}\gamma_{i,n}}{\gamma_o}\right), \label{eq:SNR_AntennaCorrOnlyFinal}
\end{align}
and $\sigma^2_{i,n}$ is the $n^{th}$ diagonal entry of $\tilde{\mathbf{R}}_i^{-1}$, i.e.
\begin{align}
\sigma^2_{i,n} = \mathbf{e}_{n,d_i}^T\tilde{\mathbf{R}}_i^{-1}\mathbf{e}_{n,d_i} \ , \label{eq:SigmaC2}
\end{align}
where $\mathbf{e}_{n,d_i}$ is the $n^{th}$ column of $\mathbf{I}_{d_i}$. Next we bound $\sigma^2_{i,n}$ using Lemma \ref{lem:BoundingSigma_in}.

\begin{lem}\label{lem:BoundingSigma_in}
Assume $\mathbf{R}_t$ is a Hermitian PSD matrix, $\mathbf{F}$ is a random $N_t\times d$ matrix such that $\mathbf{F}^*\mathbf{F}=\mathbf{I}_{d}$ and $\tilde{\mathbf{R}}=\mathbb{E}\left\{\mathbf{F}^*\mathbf{R}_t\mathbf{F}\right\}$. Then, $\sigma^2_n = \mathbf{e}_{n,d}^T\tilde{\mathbf{R}}^{-1}\mathbf{e}_{n,d}$  can be bounded as
\begin{align}
\left\{\begin{array}{ll}
\frac{1}{\lambda_N} \leq \sigma^2_{n} \leq \frac{1}{\lambda_1} &  d=1 \\
 \frac{1}{\lambda_1} + \frac{(\lambda_1 - \lambda_N)^2}{\lambda_1(\lambda_1\lambda_N -  d\parallel \mathbf{R}_t \parallel_2^2)} \leq  \sigma^2_{n} \leq \frac{1}{4\lambda_1}\left( \frac{\lambda_1}{\lambda_N} + \frac{\lambda_N}{\lambda_1} + 2 \right)& d>1 
\end{array}
\right.,\quad \forall n\in \{1,\cdots,d\}, \label{eq:BoundOnSigma_i_n}
\end{align}
where  $\lambda_1\leq\lambda_2\leq\ldots\leq\lambda_{N}$ are the eigenvalues of $\mathbf{R}_t$.
\end{lem}
\begin{IEEEproof}
The Kantrovich inequality \cite{Householder1965} states that for a $d\times d$ PSD matrix $\tilde{\mathbf{R}}$ \cite{Bai1996}
\begin{align}
\sigma^2_n  \leq \frac{1}{4\tilde{\mathbf{R}}(n,n)}\left( \frac{\lambda_{lb}}{\lambda_{ub}} + \frac{\lambda_{ub}}{\lambda_{lb}} + 2 \right)  , \label{eq:KantroSpecific_1}
\end{align}
where $\lambda_{lb}$ and $\lambda_{ub}$ are lower and upper bounds for the smallest and largest eigenvalue of $\tilde{\mathbf{R}}$ respectively.  Moreover, $ \tilde{\mathbf{R}}(n,n)=\mathbb{E}\left\{\mathbf{F}^*(n,:)\mathbf{R}_t\mathbf{F}(:,n)\right\}$ is a quadratic form and can be bounded as
\begin{align}
\lambda_1\leq\tilde{\mathbf{R}}(n,n)\leq\lambda_N , \label{eq:BoundOnDiagonalElementsOfRtDelta}
\end{align}
where $\lambda_1\leq\lambda_2\leq\ldots\leq\lambda_N$ are the eigenvalues of $\mathbf{R}_t$. In addition, since $\mathbf{F}^*\mathbf{F} =\mathbf{I}_{d}$, using Pioncare's separation theorem\cite{Lutkepohl1996}, the sorted eigenvalues of $\tilde{\mathbf{R}}$, $\tilde{\lambda}_i$ $i\in\{1,\ldots,d\}$, can be bounded as 
\begin{align}
\lambda_i \leq \tilde{\lambda}_i \leq \lambda_{N-d+i}\ \ \ i=1,\cdots,d \ . \label{eq:BoundOnEigenValuesOfRtDelta}
\end{align}
Substituting the bounds from (\ref{eq:BoundOnDiagonalElementsOfRtDelta}) and (\ref{eq:BoundOnEigenValuesOfRtDelta}) into (\ref{eq:KantroSpecific_1}) gives an upper bound on $\sigma_n^2$, i.e.
\begin{align}
\sigma_n^2 \leq \frac{1}{4\lambda_1}\left( \frac{\lambda_1}{\lambda_N} + \frac{\lambda_N}{\lambda_1} + 2 \right). \label{eq:UpperBoundOnTildeRInverse_BigOne}
\end{align}
Note that when $d=1$, (\ref{eq:CovOfDelta}) will simplify and (\ref{eq:BoundOnDiagonalElementsOfRtDelta}) gives a tighter upper-bound on $\sigma^2_{n}$. Combining the bounds from (\ref{eq:BoundOnDiagonalElementsOfRtDelta}) and (\ref{eq:UpperBoundOnTildeRInverse_BigOne}) results in
\begin{align}
\sigma_n^2 \leq \left\{ 
\begin{array}{ll} \frac{1}{4\lambda_1}\left( \frac{\lambda_1}{\lambda_N} + \frac{\lambda_N}{\lambda_1} + 2 \right)& d>1 \\
\frac{1}{\lambda_1}& d=1 \end{array} \right., \ \ \forall n\in\{1,\cdots,d\}. 
\nonumber
\end{align}

\indent As $\sum_{m=1}^{d}\parallel \tilde{\mathbf{R}}(n,m) \parallel_2^2\ \leq  d\parallel \mathbf{R}_t \parallel_2^2$ \cite{Seber2008}, by using the upper bound on the diagonal entries of $\tilde{\mathbf{R}}$ given by (\ref{eq:BoundOnDiagonalElementsOfRtDelta}), a lower bound on $\sigma^2_{n}$ can be found \cite{Robinson1992}
\begin{align}
\sigma^2_n \geq \frac{1}{\lambda_1} + \frac{(\lambda_1 - \lambda_N)^2}{\lambda_1(\lambda_1\lambda_N -  d\parallel \mathbf{R}_t \parallel_2^2)} 
\ \ \forall n\in\{1,\cdots,d\}. \label{eq:UpperBoundOnDiagonalOfInverseOfRtDelta}
\end{align}
When $d=1$, as $\parallel \mathbf{R}_t \parallel_2^2 = \lambda_N^2$, the right hand side of (\ref{eq:UpperBoundOnDiagonalOfInverseOfRtDelta}) simplifies to
\begin{align}
 \frac{1}{\lambda_1} + \frac{(\lambda_1 - \lambda_N)^2}{\lambda_1\lambda_N(\lambda_1 -  \lambda_N)} = \frac{\lambda_N(\lambda_1 -  \lambda_N) + (\lambda_1 - \lambda_N)^2}{\lambda_1\lambda_N(\lambda_1 -  \lambda_N)} = \frac{1}{\lambda_N} \ ,	\nonumber
\end{align}
which agrees with (\ref{eq:BoundOnDiagonalElementsOfRtDelta}).
\end{IEEEproof}

\indent Instead of bounding (\ref{eq:SigmaC2}), we could directly compute $\tilde{\mathbf{R}}_i$. To compute $\tilde{\mathbf{R}}_i$, correlation (covariance) function between the elements of $\mathbf{F}_i$ is needed. In the alternating minimization algorithm, the columns of $\mathbf{F}_i$ are set to the $d_i$ least dominant eigenvectors of 
\begin{align}
\sum_{k\neq i}\mathbf{H}_{k i}^*(\mathbf{I}_{N_r}-\mathbf{C}_{k}\mathbf{C}_{k}^*)\mathbf{H}_{k i} \ ,  \label{eq:PrecodForEachI}
\end{align}
and because $\mathbf{C}_{k}$ is not independent of $\mathbf{H}_{k i}$ for $i\neq k$, (\ref{eq:PrecodForEachI}) is not a Wishart matrix and the known results for the covariance matrix of the eigenvectors of Wishart matrices (e.g. see \cite{Siotani1985,Bai2007} and references therein) do not lead to an exact characterization of $\tilde{\mathbf{R}}_i$. 
Assumption \ref{Assump:MainAssump}, however, paves the way for 
directly computing (approximating) $\tilde{\mathbf{R}}_i$ using Corollary \ref{Col:PreCoders} and Lemma \ref{lem:CovarianceOfFi}.
\begin{Assump}\label{Assump:MainAssump}
Any set of basis of the interference subspace at the $k^{th}$ receiver, columns of $\mathbf{C}_k$, is independent of the channels between the interfering transmitters and the $k^{th}$ receiver, $\mathbf{H}_{ki}$ for ${i\neq k\in\{1,\ldots,K\}}$. 
\end{Assump}

Using Assumption \ref{Assump:MainAssump} and Lemma \ref{Lem:WishDist}, each term of the summation in (\ref{eq:PrecodForEachI}) is an $N_r\times N_r$ Wishart distributed matrix with $d_{k}$ degrees of freedom and covariance matrix $\mathbf{R}_t$. Moreover, as $\mathbf{H}_{ki}$ is independent of $\mathbf{H}_{nm}$ for $k\neq i$ or $i\neq m$, based on \cite[Theorem 3.4.3]{Mardia1979}, distribution of (\ref{eq:PrecodForEachI}) is given by Corollary \ref{Col:PreCoders}.

\begin{Col}\label{Col:PreCoders} In the alternating minimization algorithm, 
using Assumption \ref{Assump:MainAssump},  Lemma \ref{Lem:WishDist_Pre}, and Lemma \ref{Lem:WishDist}, (\ref{eq:PrecodForEachI}) is an $N_r\times N_r$  complex Wishart matrix with ${\sum_{k \neq i}d_{k}}$ degrees of freedom and covariance matrix $\mathbf{R}_t$. 
\end{Col}

\indent Based on Corollary \ref{Col:PreCoders}, columns of $\mathbf{F}_i$ are a subset of eigenvectors of a matrix with central complex Wishart distribution. Lemma \ref{lem:CovarianceOfFi} gives the covariance (correlation) function between components of the eigenvectors of a Wishart matrix which can be used to compute (\ref{eq:CovOfDelta}).

\begin{lem}\label{lem:CovarianceOfFi}
Let $\tilde{\lambda}_p$ and $\tilde{\bm{u}}_p$, for $p=1,\ldots,N$, be the eigenvalues and the eigenvectors of an $N\times N$ matrix with a central complex Wishart distribution with $D$ degrees of freedom and covariance matrix $\mathbf{R}_t$. Next, let $\lambda_p$ and $\pmb{u}_p$, for $p=1,\ldots,N$, be the eigenvalues and eigenvectors of $\mathbf{R}_t$. Also let the $q^{th}$ elements of $\tilde{\bm{u}}_p$ and $\boldsymbol{u}_p$ be $\tilde{u}_{pq}$ and $u_{pq}$ respectively. Assuming $\lambda_1>\lambda_2>\cdots>\lambda_{N}>0$, for $D\rightarrow\infty$ and fixed $N$
\begin{align}
\mathbb{E}\left\{\tilde{\bm{u}}_p\right\} &= \boldsymbol{u}_p\ ,  \label{eq:MeanOfFiElements} \\
\rm{cov}(\tilde{u}_{pq},\tilde{u}_{p\prime q \prime}) &= \left\{ \begin{array}{ll}
\cfrac{\lambda_p}{D}\displaystyle\sum_{r=1,r\neq p}^{N}\cfrac{\lambda_r u_{rq} u_{rq\prime}^*}{(\lambda_r-\lambda_p)^2}\  \quad & p=p\prime 
\\
-\cfrac{\lambda_p\lambda_{p\prime} u_{pq} u_{p\prime q \prime}^*}{D(\lambda_p-\lambda_{p\prime})^2}\  \quad & p \neq p\prime 
\end{array}
\right. \ . \label{eq:CovBetweenFiElements}
\end{align}
\end{lem}
\begin{IEEEproof}
See \cite[Chap. 3.6]{Jolliffe2002}, \cite[Chap. 10]{Siotani1985} and references therein.
\end{IEEEproof}
Using Lemma \ref{lem:CovarianceOfFi}, $\tilde{\mathbf{R}}_i(n,m)\  \forall n,m\in \{1,\ldots,d_i\}$ in (\ref{eq:CovOfDelta}) is given by
\begin{align}
\tilde{\mathbf{R}}_i(n,m)
&= \rm{tr}\left(\mathbb{E}\left\{\mathbf{F}_i^*(n,:)\mathbf{R}_t\mathbf{F}_i(:,m)\right\}\right) = \rm{tr}\left(\mathbb{E}\ \{\mathbf{F}_i(:,m)\mathbf{F}_i^*(n,:)\}\mathbf{R}_t\right) \nonumber \\
 &= \rm{tr}\left( \left( \rm{cov}\left(\mathbf{F}_i(:,m),
\mathbf{F}_i(:,n)\right)+ 
\mathbb{E}\left\{\mathbf{F}_i(:,m)\right\}\mathbb{E}\left\{\mathbf{F}_i ^*(n,:)\right\}  
\right)\mathbf{R}_t\right). \label{eq:FormulaForRiBasedOnCov}
\end{align}
Moreover, values given by (\ref{eq:FormulaForRiBasedOnCov}) should be normalized by noting that $\mathbf{F}_i^*\mathbf{F}_i=\mathbf{I}_{d_i}$ and hence
\begin{align}
\rm{tr}\left(  \rm{cov}\left(\mathbf{F}_i(:,m),
\mathbf{F}_i(:,m)\right)+ 
\mathbb{E}\left\{\mathbf{F}_i(:,m)\right\}\mathbb{E}\left\{\mathbf{F}^*_i(m,:)\right\}  
\right) = 1 \quad \forall m\in\{1,\ldots,d_i\}. \nonumber
\end{align}

\indent Note that the asymptotic expressions presented in Lemma \ref{lem:CovarianceOfFi} are valid when the number of streams in the IA network is asymptotically large while the number of antennas in each node is kept constant. We know, however, for a K-user interference channel, when the number of streams in the network increases, IA remains feasible only if the number of antennas at each node increases accordingly\cite{Yetis2009}. 
But such asymptotic results fall into the category of asymptotic behavior of eigenvectors of large covariance matrices (e.g. see \cite{Bai2007} and references therein) where closed form solutions similar to (\ref{eq:CovBetweenFiElements}) are either not available or their complexity will not benefit the current discussion of this paper. It should be noted that more accurate results for the values of $\tilde{\mathbf{R}}_i$ can always be obtained by using a better approximation or a more complex closed-form expression.

\indent Under certain models of spatial correlation, for example those based on scattering clusters \cite{Spencer2000}, the spatial correlation is a function of cluster locations and will vary for different transmit and receiver locations. 
Now we assume there is a potentially different transmit correlation for each link pair, i.e. $\mathbf{R}_{t_{i}}\neq \mathbf{R}_{t_{k}}$ for $k\neq i\in\{1,\ldots,K\}$. Using Assumption \ref{Assump:MainAssump}, (\ref{eq:PrecodForEachI}) will be a linear sum of Wishart matrices with equal degrees of freedom but unequal covariance matrices. It is shown in \cite{Nel1986} that such a linear sum can be approximated with another Wishart matrix whose degrees of freedom and covariance matrix are given by Lemma \ref{lem:SumOfWish}.

\begin{lem}\label{lem:SumOfWish}
Suppose $\bar{\mathbf{T}} = \sum_{i=1}^{K-1}\mathbf{T}_i$ where $\mathbf{T}_i$ is a $N\times N$ Wishart distributed matrix with $d_i$ degrees of freedom and covariance matrix $\mathbf{R}_{t_i}$. Then $\bar{\mathbf{T}}$ has, approximately, Wishart distribution with $\bar{d}$ degrees of freedom and covariance matrix $\bar{\mathbf{R}}_t$ where
$\bar{d} = \frac{\rm{tr}\left(\sum_{i=1}^{K-1}d_i\mathbf{R}_{t_i}\right)^2 +\rm{tr}^2\left(\sum_{i=1}^{K-1}d_i\mathbf{R}_{t_i}\right)}
{\sum_{i=1}^{K-1}d_i\left(\rm{tr}(\mathbf{R}_{t_i}^2)+\rm{tr}^2(\mathbf{R}_{t_i})\right)}$ and $\bar{\mathbf{R}}_t = \frac{1}{\bar{d}}\sum_{i=1}^{K-1}d_i\mathbf{R}_{t_i}$.
\end{lem}
\begin{IEEEproof}
See \cite[Section 3]{Nel1986}
\end{IEEEproof}
Using Lemma \ref{lem:CovarianceOfFi} and Lemma \ref{lem:SumOfWish}, the correlation between the elements of eigenvectors of (\ref{eq:PrecodForEachI}) can be found and consequently one can compute (\ref{eq:CovOfDelta}) for each receiver. Note that if $\mathbf{R}_{t_i}\neq \mathbf{R}_{t_k}$ for $i\neq k\in\{1,\ldots,K\}$, then $\sigma^2_{i,n}\neq \sigma^2_{k,n}$. 
For the rest of this manuscript, to simplify the notation, we assume equal transmit correlations at each transmitter. The forthcoming analysis and the derived equations, however, can be generalized to unequal transmit correlation matrices.

\section{Imperfect Channel Knowledge} \label{sec:IPCSI}
The channel state is estimated and thus known imperfectly in realistic wireless systems. 
We model imperfect CSI through a Gauss-Markov uncertainty of the form \cite{Wang2007,Musavian2007}
\begin{align}
\mathbf{H}^w = \sqrt{1-\beta^2}\hat{\mathbf{H}}^{w} + \beta\mathbf{E} \ , \label{eq:ImperfectChannelStateInformationGeneralForm}
\end{align}
where $\mathbf{H}^w\sim \mathcal{CN}(\mathbf{0},\mathbf{I})$ is the true Gaussian part of the channel matrix, $\hat{\mathbf{H}}^w\sim \mathcal{CN}(\mathbf{0},\mathbf{I})$ is the imperfect observation of $\mathbf{H}^w$ available to the nodes and $\mathbf{E}\sim \mathcal{CN}(\mathbf{0},\mathbf{I})$ is an i.i.d Gaussian noise term.
The parameter $\beta$ characterizes the partial CSI since $\beta =0$ corresponds to perfect channel knowledge and $\beta= 1$ corresponds to no CSI knowledge and values of $0<\beta<1$ account for partial CSI. Note that our forthcoming discussion on the general expression of the imperfect CSI at the nodes as modeled in (\ref{eq:ImperfectChannelStateInformationGeneralForm}) can be used to study specific scenarios of imperfect CSI such as channel estimation error or analog feedback. The difference between such scenarios is how $\beta$ changes as a function of different system parameters. For example, with MMSE channel estimation, $\beta$ is a function of pilot symbol SNR \cite{JindalMIMOtraining2010} or for an analog feedback link, $\beta$ is a function of the number of channel uses per channel coefficient and the SNR of the feedback link \cite{JindalAnalogFeedback2007}. We assume $\beta$ is a constant but one can extend the results to $\beta$ being an arbitrary function of the system parameters.

\indent  In presence of a transmit correlation, $\mathbf{R}_t$, using arguments similar to \cite{Ding2009,Yoo2004}, we assume that the transmit correlation varies slower than the channel itself such that the nodes have a perfect estimate of $\mathbf{R}_t$ even if their observation of $\mathbf{H}$ is not perfect. Therefore, the channel correlation given by (\ref{eq:AntennaCorrModel}) can be considered in conjunction with the CSI imperfection using the following model
\begin{align}
\mathbf{H} = (\sqrt{1-\beta^2}\hat{\mathbf{H}}^{w} + \beta\mathbf{E})\mathbf{R}_t^{1/2}  . \label{eq:IPCSI_Model}
\end{align}

\indent We assume that the precoding and combining matrices are designed using $\hat{\mathbf{H}}^w\mathbf{R}_t^{1/2}$, thus effectively ignoring the introduced CSI imperfection. Similar to (\ref{eq:RxSignal_origin}) and (\ref{eq:yiHeff})
\begin{align}
&\mathbf{y}_i =  \sum_{k=1}^{K}\left(\sqrt{1-\beta^2}\hat{\mathbf{H}}_{ik}^w + \beta \mathbf{E}_{ik}\right)\mathbf{R}_t^{1/2}\mathbf{F}_{k}\mathbf{x}_{k} + \mathbf{z}_i \nonumber \\
& = \sqrt{1-\beta^2}\hat{\mathbf{H}}_i
\left[\begin{array}{cc}
\mathbf{x}_i \\
\sum_{k\neq i}^K\hat{\mathbf{A}}_{ik}\mathbf{x}_{k} \end{array}\right] + \beta\mathbf{E}_i(\mathbf{R}_t^{1/2}\otimes \mathbf{I}_K)\mathbf{F}\mathbf{x} + \mathbf{z}_i \ , \label{eq:ReceivedWithCU}
\end{align}
where $\hat{\mathbf{H}}_i = [\hat{\mathbf{H}}_{ii}^w\mathbf{R}_t^{1/2}\mathbf{F}_i,\mathbf{C}_i]$, $\hat{\mathbf{A}}_{ik}$ is found by solving $\mathbf{C}_i\hat{\mathbf{A}}_{ik} = \hat{\mathbf{H}}_{ik}^w\mathbf{R}_t^{1/2}\mathbf{F}_{k}$, $\mathbf{F}$ and $\mathbf{E}_i$ are block diagonal matrices with $\mathbf{F}_1,\ldots,\mathbf{F}_K$ and $\mathbf{E}_{i1},\ldots,\mathbf{E}_{iK}$ on their main diagonals respectively and ${\mathbf{x} = \big[\mathbf{x}_i^T,\ldots,\mathbf{x}_K^T]^T}$. The ZF equalizer, similar to (\ref{eq:ZFpcsi}), is given by $\mathbf{B}_i\hat{\mathbf{H}}_i^{-1}$ and the distribution of the post-processing SINR after applying the ZF equalizer to (\ref{eq:ReceivedWithCU}) is given by Lemma \ref{lem:ImperfectCSI_Correlation_ReceivedSINRdistributino}.

\begin{lem}\label{lem:ImperfectCSI_Correlation_ReceivedSINRdistributino}
Given the channel model of (\ref{eq:IPCSI_Model}), assuming the IA precoders are designed ignoring the CSI imperfection, the post-processing SINR distribution after applying a ZF equalizer is given by
\begin{align}
f^{ci}(\gamma_{i,n}) =  1/\alpha_i
\exp\left(-\gamma_{i,n}/\alpha_i\right) \quad i\in\{1,\ldots,K\}, n\in\{1,\ldots,d_i\}, 
\label{eq:pdfSNRwithErrorAndCorr}
\end{align}
where $\alpha_i = \frac{1-\beta^2}{\sigma^2_{i,n}d_i(\beta^2\mathcal{I}+\frac{1}{\gamma_o})}$, $\sigma^2_{i,n}$ is given by (\ref{eq:SigmaC2}) and $\mathcal{I} = \sum_{i=1}^{K}\rm{tr}\big(\mathbf{F}_i^*\mathbf{R}_t\mathbf{F}_i\big)/d_i$.
\end{lem}
\begin{IEEEproof}
Applying the ZF equalizer to  (\ref{eq:ReceivedWithCU}), the per stream SINR of the $i^{th}$ user is given by
\begin{align}
&\gamma_{i,n}^{ci} = \frac{(1-\beta^2)/d_i}
{\left[ \mathbf{B}_i\left(\beta^2\hat{\mathbf{H}}_i^{-1}
\mathbf{E}_i(\mathbf{R}_t^{1/2}\otimes \mathbf{I}_K)\mathbf{F}\mathbf{D}\mathbf{F}^*(\mathbf{R}_t^{1/2}\otimes \mathbf{I}_K)^*\mathbf{E}_i^*\big(\hat{\mathbf{H}}_i^{-1}\big)^*
 + \frac{1}{\gamma_o}\big( \hat{\mathbf{H}}_i^*\hat{\mathbf{H}}_i\big)^{-1} \right) \mathbf{B}_i^* \right]_{n,n}}\ , \label{eq:SINR_temp1}
\end{align}
where ${\mathbf{D} = \mathbb{E}\left\{\mathbf{x}\mathbf{x}^*\right\}}$. As $\mathbf{E}_i$ has i.i.d unit-variance zero-mean terms, we have $\mathbb{E}\left\{\mathbf{A} \mathbf{E}_i \mathbf{B}\mathbf{E}_i^*\mathbf{C}\right\} = \rm{tr}(\mathbf{B})\mathbf{A}\mathbf{C}$ for any matrices $\mathbf{A}$,$\mathbf{B}$ and $\mathbf{C}$ independent of $\mathbf{E}_i$ \cite[Section 21.2]{Seber2008}. Therefore, (\ref{eq:SINR_temp1}) simplifies to
\begin{align}
\gamma_{i,n}^{ci}&= \frac{(1-\beta^2)/d_i}
{\left[\mathbf{B}_i\left( \beta^2\rm{tr}\big((\mathbf{R}_t^{1/2}\otimes \mathbf{I}_K)\mathbf{F}\mathbf{D}\mathbf{F}^*(\mathbf{R}_t^{1/2}\otimes \mathbf{I}_K)^*\big)\hat{\mathbf{H}}_i^{-1}\big(\hat{\mathbf{H}}_i^{-1}\big)^*
 + \frac{1}{\gamma_o}\big( \hat{\mathbf{H}}_i^*\hat{\mathbf{H}}_i\big)^{-1} \right)\mathbf{B}_i^*  \right]_{n,n}} \nonumber
\\
&=\frac{(1-\beta^2)/d_i}
{\left[\mathbf{B}_i\left(\beta^2\rm{tr}\big((\mathbf{R}_t\otimes \mathbf{I}_K)\mathbf{F}\mathbf{D}\mathbf{F}^*\big)\big( \hat{\mathbf{H}}_i^*\hat{\mathbf{H}}_i\big)^{-1}
 + \frac{1}{\gamma_o}\big( \hat{\mathbf{H}}_i^*\hat{\mathbf{H}}_i\big)^{-1} \right)\mathbf{B}_i^*  \right]_{n,n}} \label{eq:temp11} , \
\end{align}
where the equality ${\rm{tr}(\mathbf{A}\mathbf{B}\big)= \rm{tr}\big(\mathbf{B}\mathbf{A})}$ was used in (\ref{eq:temp11}). Exploiting the block diagonal structure of $\mathbf{F}$ and $(\mathbf{R}_t\otimes \mathbf{I}_K)$ results in
\begin{align} 
\gamma_{i,n}^{ci} 
&=  \frac{(1-\beta^2)/d_i}
{\left[\mathbf{B}_i\left( (\beta^2\mathcal{I}+\frac{1}{\gamma_o})\big( \hat{\mathbf{H}}_i^*\hat{\mathbf{H}}_i\big)^{-1}
 \right)\mathbf{B}_i^*  \right]_{n,n}}. 
\label{eq:SNR_withErrorAndCorr}
\end{align} 
Similar to (\ref{eq:SNR_NoError}), (\ref{eq:SNR_withErrorAndCorr}) further simplifies to 
\begin{align}
\gamma_{i,n}^{ci} = \frac{(1-\beta^2)/d_i}
{(\beta^2\mathcal{I}+1/\gamma_o)\left[\left(\hat{\mathbf{\Delta}}_i
(\mathbf{I}_{N_r}-\mathbf{C}_i\mathbf{C}_i^*)\hat{\mathbf{\Delta}}_i^*\right)^{-1}\right]_{n,n}} \ , \label{eq:SNR_Err_Corr_temp1}
\end{align}
where $\hat{\mathbf{\Delta}}_i = \mathbf{F}_i^*(\mathbf{R}_t^{1/2})^*(\hat{\mathbf{H}}_{ii}^w)^*$. The SINR distribution follows from comparing (\ref{eq:SNR_Err_Corr_temp1}) to (\ref{eq:SNR_NoErrorExpanded_AntennaCorr}).
\end{IEEEproof}
For $\mathbf{R}_t \neq \mathbf{I}$, $\mathcal{I}$ can either be approximated using Lemma \ref{lem:CovarianceOfFi} or bounded using (\ref{eq:BoundOnSigma_i_n}) by summing over the bounds of diagonal entries of $\tilde{\mathbf{R}}_i$. Note that for $\mathbf{R}_t = \mathbf{I}$, the values of $\mathcal{I}$ and $\sigma^2_{i,n}$ in (\ref{eq:SNR_withErrorAndCorr}) and (\ref{eq:SigmaC2}) will be exact and equal to $K$ and $1$ respectively. Moreover, as expected, for $\beta=0$ and $\sigma^2_{i,n}=1$, (\ref{eq:pdfSNRwithErrorAndCorr}) reduces to (\ref{eq:pdfNoError}). Comparing (\ref{eq:pdfNoError}) to (\ref{eq:pdfSNRwithErrorAndCorr}), the mean SINR at each stream has reduced by a factor of ${\sigma^2_{i,n}(\gamma_o\beta^2\mathcal{I}+1)/(1-\beta^2)}$ which results in the mean post-processing SINR reaching the maximum value of $(1-\beta^2)/(\sigma^2_{i,n}d_i\beta^2\mathcal{I})$ as $\gamma_o\rightarrow \infty$, implying a symbol-error-rate (SER) floor and a sum rate cap given $\beta\neq 0$. Note that when $\beta=0$, the mean SINR still scales linearly with increasing $\gamma_o$ and the effect of antenna correlation can be seen as a shift in SER or sum rate curves. Moreover, if $\beta$ decreases with increasing $\gamma_o$, provided that $\mathbf{R}_t$ is not rank deficient, there will be no SER floor or sum rate cap and the full multiplexing gain promised by IA will be attainable.

\section{Comparison with Point-to-Point MIMO}\label{sec:P2P}
In an interference channel, instead of utilizing IA, nodes can access the network resources in an orthogonal fashion (TDMA/FDMA). In the resulting parallel point-to-point links, nodes can apply any of the traditional MIMO transmission strategies. In this section, we discuss the two most common strategies, beamforming and SM, and present the post-processing SNR distributions in each case. The post-processing SNR distributions can be used to compare IA to beamforming and SM for a wide range of system performance measures. For the point-to-point MIMO links, we assume existence of transmit correlation and imperfect CSI similar to (\ref{eq:IPCSI_Model}).

\indent A transmitter with CSI can use beamforming to send a single stream to its receiver. Assume the transmitter ignores the channel imperfection in (\ref{eq:IPCSI_Model}) and treats $\hat{\mathbf{H}}^w\mathbf{R}_t^{1/2}$ as the true channel. To maximize the received SNR, the precoding and combining vectors are set to ${\mathbf{V}(:,1)}$ and ${\mathbf{U}(:,1)^*}$, respectively, where $\mathbf{U}$ and $\mathbf{V}$ are given by the singular value decomposition of $\hat{\mathbf{H}}^w\mathbf{R}_t^{1/2} = \mathbf{U}\boldsymbol{\Sigma}\mathbf{V}^*$ and $\bm{\Sigma}$ is diagonal matrix holding the singular values of $\hat{\mathbf{H}}^w\mathbf{R}_t^{1/2}$ in descending order. The post-processing SNR is then
\begin{align}
\gamma^{BF} = \frac{(1-\beta^2)\hat{\lambda}_1}{\beta^2
\nu
 + 1/\gamma_o} \ , \ \label{eq:Distribution_P2P_BF}
\end{align}
where $\nu = \mathbb{E}\left\{\rm{tr}\left(\mathbf{V}^*(:,1)\mathbf{R}_t\mathbf{V}(:,1)\right)\right\}$ and  $\hat{\lambda}_1$ is the largest eigenvalue of $\hat{\mathbf{H}}^w\mathbf{R}_t^{1/2}(\mathbf{R}_t^{1/2})^*(\hat{\mathbf{H}}^w)^*$ with correlated central complex Wishart distribution.  Note that columns of $\mathbf{V}$  are also the eigenvectors of $(\mathbf{R}_t^{1/2})^*(\hat{\mathbf{H}}^w)^*\hat{\mathbf{H}}^w\mathbf{R}_t^{1/2}$ which has central complex Wishart distribution with covariance matrix $\mathbf{R}_t$ and $N$ degrees of freedom and similar to (\ref{eq:CovOfDelta}), Lemma \ref{lem:CovarianceOfFi} can be used to compute $\nu$. Therefore, the distribution of the beamforming SNR is given by
\begin{align}
f^{BF}(\gamma) = 
\frac{(1-\beta^2)f^{cw}(\mathbf{R}_t,\gamma)}{\beta^2\nu + 1/\gamma_o},
\end{align}
where $f^{cw}(\mathbf{R}_t,\gamma)$ is the distribution of the largest eigenvalue of a correlated central complex Wishart matrix with covariance matrix $\mathbf{R}_t$\cite[Section IV]{Zanella2008}. 

\indent In absence of transmit CSI, SM can be used to convey $N$ data streams to the receiver. Assume the channel imperfection is ignored at the receivers and the ZF equalizer is given by  $(\hat{\mathbf{H}}^w\mathbf{R}_t^{1/2})^{-1}$. The post-processing SINR per stream can be written as 
\begin{align}
\gamma^{SM}_{n} = \frac{(1-\beta^2)/N}{ \left(\beta^2\rm{tr}(\mathbf{R}_t)/N + \frac{1}{\gamma_o}\right)
 \left[ \left((\mathbf{R}_t^{1/2})^*(\mathbf{H}^w)^*\mathbf{H}^w\mathbf{R}_t^{1/2}
\right)^{-1}\right]_{n,n}} \ , \nonumber
\end{align}
which is distributed as
\begin{align}
f_n^{SM}(\gamma_n) = \frac{1}{\omega}\exp{\left(\frac{-\gamma_n}{\omega}\right)} \ , \label{eq:SNR_P2P}
\end{align}
where $\omega = \frac{1-\beta^2}{\sigma_{n}^2(n)\left(\beta^2\rm{tr}(\mathbf{R}_t) + \frac{N}{\gamma_o}\right)}$ and $\sigma_{n}^2(n)$ is the $n^{th}$ diagonal entry of $\mathbf{R}_t^{-1}$.

\indent Using (\ref{eq:SNR_withErrorAndCorr}), (\ref{eq:Distribution_P2P_BF}) and (\ref{eq:SNR_P2P}), the performance of beamforming, SM and IA can be compared for a wide range of system performance metrics. One such metric is the achievable throughput given by
\begin{align}
R_{\rm{sum}}(\gamma_o,\beta,\alpha) = \sum_{n=1}^{\hat{d}}\int_{0}^{\infty}\log_2(1+\gamma)f(\gamma)d\gamma, \label{eq:SumRateGeneral}
\end{align}
where $\hat{d}$ is the number of streams in the network equal to $\sum_i d_i$, $N$ and $1$ for IA, SM and beamforming, respectively. Another point of comparison between IA and point-to-point MIMO links is the per-stream SER. For any modulation $\mathcal{M}$, the AWGN SER is a function of $\gamma$, e.g. $\mathcal{E}_{\mathcal{M}}(\gamma)$, and the per-stream SER can be written as
\begin{align}
\rm{SER}_{\mathcal{M}}(\gamma_o,\beta,\alpha) = \int_{0}^{\infty}\mathcal{E}_{\mathcal{M}}(\gamma)f(\gamma)d\gamma, \nonumber
\end{align}
where $f$ is any of the distributions given by (\ref{eq:SNR_withErrorAndCorr}), (\ref{eq:Distribution_P2P_BF}) or (\ref{eq:SNR_P2P}). Note that both (\ref{eq:SNR_withErrorAndCorr}) and (\ref{eq:SNR_P2P}) are exponentially distributed and the mean value of the distributions suffices to compare IA and SM in terms of per-stream SER (assuming $d_i=d_j \ \forall i,j\in\{ 1,\ldots,K\}$). Specifically, the ratio of the mean SINR of the SM link to the mean SINR of the IA network is given by
\begin{align}
\frac{\mathbb{E}\left\{f^{SM} \right\}}{\mathbb{E}\left \{f^{ci} \right\}} = \frac{ \sigma^2_{i,n}d \left(\beta^2\mathcal{I} + \frac{1}{\gamma_o}\right)}{\sigma_{n}^2(n)\left(\beta^2 + \frac{N}{\gamma_o}\right)}, \label{eq:RatioOfP2PtoIA}
\end{align}
where $\sigma^2_{i,n}=\sigma^2_{j,n}$ for $i,j\in\{1,\ldots,K\}$.
When (\ref{eq:RatioOfP2PtoIA}) results in a value greater than $1$, the (per stream) mean SINR of the SM MIMO network is higher, i.e. when (\ref{eq:RatioOfP2PtoIA}) is larger than $1$, given a SER constraint, a point-to-point SM MIMO link will satisfy that constraint with a smaller transmit power. 

\section{Numerical Results} \label{Sec:Numerical}
Consider a $K$-user $N\times N$ MIMO interference channel. All the channel and noise coefficients are distributed as $\mathcal{CN}(0,1)$.
To focus on the distribution of the received SINR, we exclude any large-scale fading effects from the channel model effectively assuming all the nodes are co-located and equipped with omni-directional antennas. It is shown in \cite{Peters2009} that using IA, each transmitter can send a single stream to its corresponding receiver as long as $N = \lceil(1+K)/2\rceil$. In the forthcoming discussions, the channel coefficient matrices are normalized such that $\parallel \mathbf{H} \parallel_F^2 = N_rN_t$ which given the transmit correlation and channel imperfection model of (\ref{eq:ImperfectChannelStateInformationGeneralForm}) translates to $\rm{tr}\left( \mathbf{R}_t \right) = N_t$. Note that there exist constraints on the off-diagonal elements of the $\mathbf{R}_t$ due to the transmit correlation matrix being Hermitian PSD. In this section, for the transmit correlation matrix, we adapt a $N\times N$ transmit correlation matrix of the form $\mathbf{R}_t(i,j) = \alpha^{|i-j|}$ for $i,j\in\{1,\cdots,N\}$, where $\alpha\in \mathbb{C}\ $ is such that $\mathbf{R}_t$ is positive definite. This model, widely used in literature and industry \cite{R4_060101}, models the correlation between elements of a uniform linear array antenna where $\alpha=0$ and $|\alpha|=1$ correspond to no correlation and a rank $1$ channel, respectively. 

\begin{figure}[t]
 \centering
\includegraphics[bb= 0 0 350 290 ,width=3.5in]{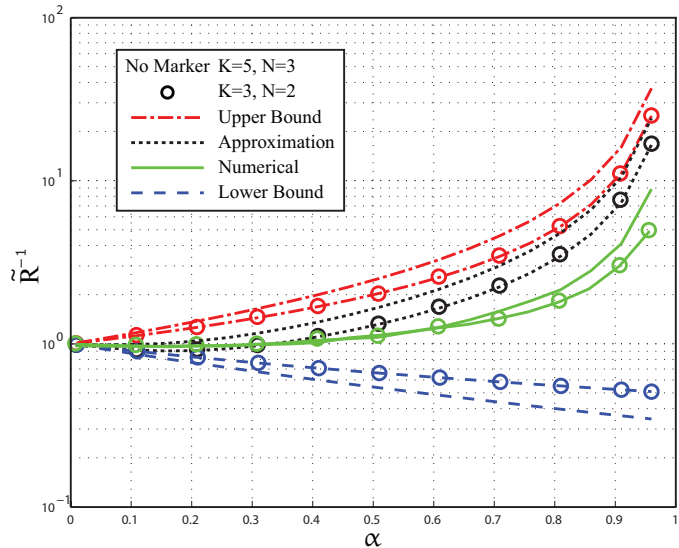}
\caption{Comparison of the numerical values of $\tilde{R}_i^{-1}$ with the proposed approximation and the bounds in Section \ref{sec:Tcorr} in two MIMO IA networks, a $3$-user $2\times 2$ and a $5$-user $3\times 3$, for varying antenna correlation values. As expected, the approximation is closer to the true value compared to the bounds. The proposed approximation is within $10\%$ of the true value for $\alpha<0.3$.}
\label{fig:RtDelta}
\end{figure}

\indent The numerical values of $\tilde{R}^{-1}_i$ computed from (\ref{eq:CovOfDelta}), the theoretical approximation found by replacing (\ref{eq:CovBetweenFiElements}) and (\ref{eq:MeanOfFiElements}) into (\ref{eq:CovOfDelta}) and the bounds given by (\ref{eq:BoundOnSigma_i_n}) versus $\alpha$ for two different IA networks are depicted in Fig. \ref{fig:RtDelta}. The two IA networks are a $3$-user $2\times 2$ and a $5$-user $3\times 3$ MIMO networks with $d_i=1, i\in\{1,\ldots,K\}$. Both networks use the alternating minimization algorithm to compute the precoding and combining matrices. As expected, the theoretical approximation is a better estimate of the true values of $\tilde{R}_i^{-1}$ than the presented upper and lower bounds, especially for small values of $\alpha$. Note that by increasing $N$ and $\alpha$, the accuracy of the approximation decreases.  

\indent In computing (\ref{eq:pdfSNRwithErrorAndCorr}), for a fixed $\gamma_o$, as $\beta$ increases, the numerator approaches zero and the impact of any errors in approximating $\sigma^2_{i,n}$ on (\ref{eq:pdfSNRwithErrorAndCorr}) diminishes. Therefore, we expect that by increasing $\beta$ the distribution of the received SINR given by (\ref{eq:pdfSNRwithErrorAndCorr}) will approach to the true distribution. Moreover, for a fixed $\beta$, as $1/\gamma_o$ approaches zero, any errors in computing $\sigma^2_{i,n}$ will be attenuated by a smaller value and we expect (\ref{eq:pdfSNRwithErrorAndCorr}) to become closer to the true value of the distribution. Although there exists numerous methods of quantifying the accuracy of (\ref{eq:pdfSNRwithErrorAndCorr}), we choose the Kullback-Leibler divergence (KLD) \cite{KLD1951} between the distribution given by (\ref{eq:pdfSNRwithErrorAndCorr}) and the empirical one. The continuous distributions required to compute KLD are approximated with uniformly spaced samples in their range.

\begin{figure}[t]
 \centering
\includegraphics[bb= 0 0 370 290,width=4.1in]{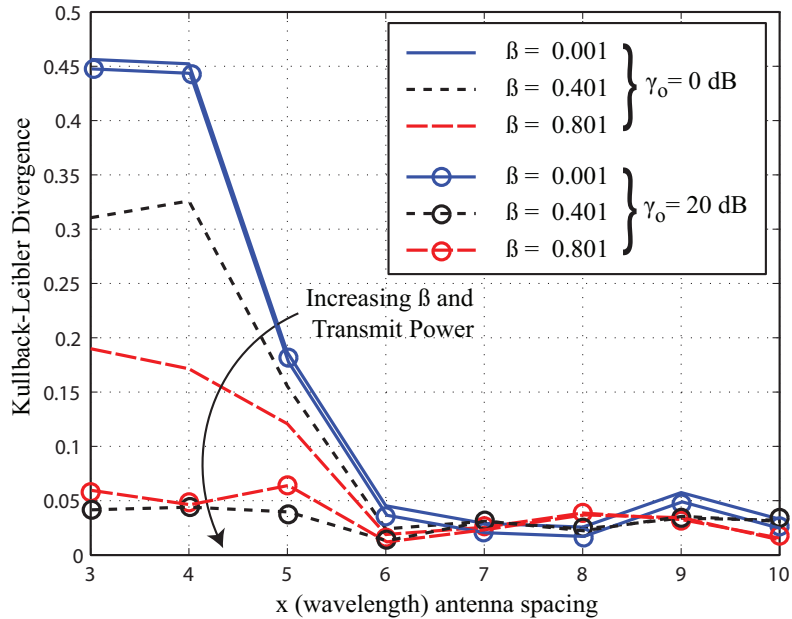}
\caption{Kullback-Leibler divergence between the theoretical distribution of the per-stream received SINR given by (\ref{eq:pdfSNRwithErrorAndCorr}) and the true distribution found through numerical simulation as a function of $\alpha$ for varying $\beta$ and $\gamma_o$. As can be seen, the accuracy of (\ref{eq:pdfSNRwithErrorAndCorr}) increases with increasing $\beta$ and transmit power.}
\label{fig:KLD}
\end{figure}

\indent Now consider a $3$-user IA network with $d_i=1 \ \forall i\in\{1,2,3\}$ and transmit nodes equipped with a uniform linear array of $2$ antennas spaced $a\lambda$ apart where $a\in \mathbb{R}^+$ and $\lambda$ is the transmission signal wavelength. Here, we control $\alpha$ by varying $a$. We use the method proposed in \cite{R4_060101} (\textit{Suburban Macro-cell} environment) to calculate $\alpha$ for various antenna spacings. Table \ref{table:AlphaValues} shows the selected antenna spacings and the corresponding $\alpha$ values. The KLD between empirical distribution of the received SINR and theoretical distribution given by (\ref{eq:pdfSNRwithErrorAndCorr}) as a function of $\alpha$ for three values of $\beta$ and two values of $\gamma_o$ is shown in Fig. \ref{fig:KLD}. As expected, the divergence between the distributions decreases by increasing $\beta$ and $\gamma_o$. 

\begin{figure}[t]
 \centering
\includegraphics[bb= 0 0 400 290,width=4.1in]{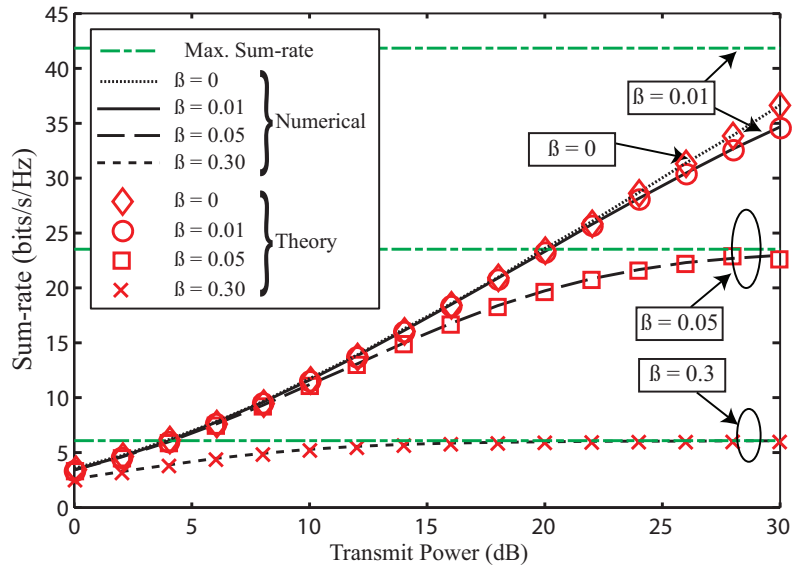}
\caption{Theoretical predictions and numerical results for the sum rate of a $4$-user $3\times 3$ constant channel MIMO IA ($\alpha=0$) for $4$ values of $\beta$ versus $\gamma_o$. As can be seen, numerical simulations closely follow the analytical curves and the theoretical maximum sum rates correctly predict the upper bounds for the given $\beta$.}
\label{fig:Sumrate}
\end{figure}

\indent As discussed in Section \ref{sec:IPCSI}, the distribution of per stream SINR given by (\ref{eq:pdfSNRwithErrorAndCorr}) is exact for $\alpha=0$. 
The sum rate of a $4$-user $3\times 3$ MIMO system, when $\mathbf{R}_t=\mathbf{I}$, for four values of $\beta$ versus $\gamma_o$ is depicted in Fig. \ref{fig:Sumrate} where theoretical sum rate curves are found by replacing (\ref{eq:pdfSNRwithErrorAndCorr}) into (\ref{eq:SumRateGeneral}) and the maximum achievable sum rates (when $\beta>0$) are found by replacing the mean value of (\ref{eq:pdfSNRwithErrorAndCorr}) with its limit as $\gamma_o$ approaches infinity, i.e. $\frac{1-\beta^2}{\sigma^2_{i,n}d_i\beta^2\mathcal{I}}$. As can be seen, the numerical results exactly follow the theoretical predictions. Moreover, when $\beta\neq 0$ the multiplexing gain is zero. For $\beta\neq 0$, however, by comparing the sum rate curve with the case of perfect CSI ($\beta=0$), one could find ranges of transmit power where channel imperfection has practically no effect on the sum rate; for example, $\gamma_o<20dB$ for $\beta=0.01$. 

\begin{figure}[t]
 \centering
\includegraphics[bb= 0 0 370 290,width=3.75in]{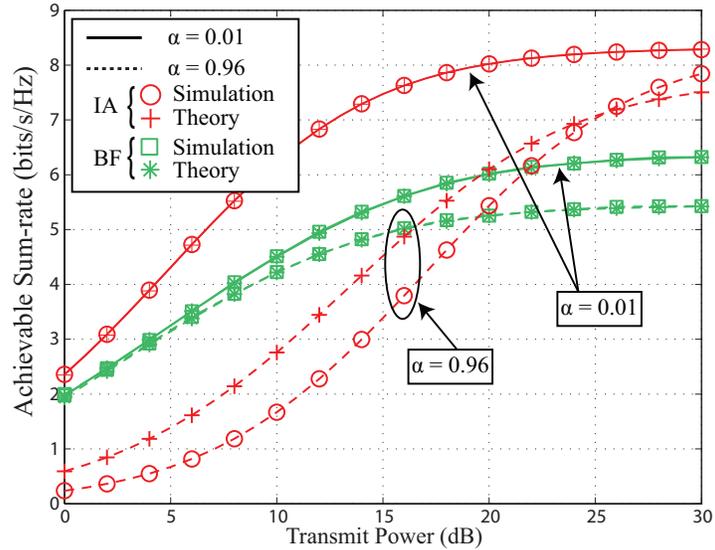}
\caption{Achievable sum rate of $3$-user $2\times 2$ IA MIMO network and a $2\times 2$ MIMO beamforming link as a function of $\gamma_o$ for varying $\alpha$ and a fixed $\beta=0.19$. The accuracy of (\ref{eq:pdfSNRwithErrorAndCorr}) increases with increasing $\gamma_o$ and decreases with increasing $\alpha$. As can be seen, beamforming can outperform IA for a wide range of parameters. Moreover, Lemma \ref{lem:CovarianceOfFi} can be used to accurately predict the beamforming performance.}
\label{fig:IA_BF_SR}
\end{figure}

\indent The effect of transmit correlation and imperfect CSI on the sum rates of a $3$-user $2\times 2$ MIMO IA network and a $2\times 2$ point-to-point MIMO beamforming link is shown in Fig. \ref{fig:IA_BF_SR}. Several important conclusions can be drawn from this figure. First, for non-ideal transmit correlation and imperfect CSI, sum rate of a beamforming link can be higher than the sum rate of an IA network with the cross-over point between the sum rate curves being a function of $\alpha$, $\beta$ and $\gamma_o$. Second, the accuracy of (\ref{eq:pdfSNRwithErrorAndCorr}), as discussed before, increases with increasing $\gamma_o$ while it decreases with increasing $\alpha$. Note that as $\alpha$ increases, $\mathbf{R}_t$ approaches a rank-deficient matrix (violating our assumptions in Lemma \ref{lem:CovarianceOfFi}) which reduces the accuracy of (\ref{eq:pdfSNRwithErrorAndCorr}). Third, Lemma \ref{lem:CovarianceOfFi} can be used to accurately predict the performance of beamforming in presence of transmit correlation and channel imperfection.

\begin{figure}[t]
 \centering
\includegraphics[bb= 0 0 410 360 ,width=3.75 in]{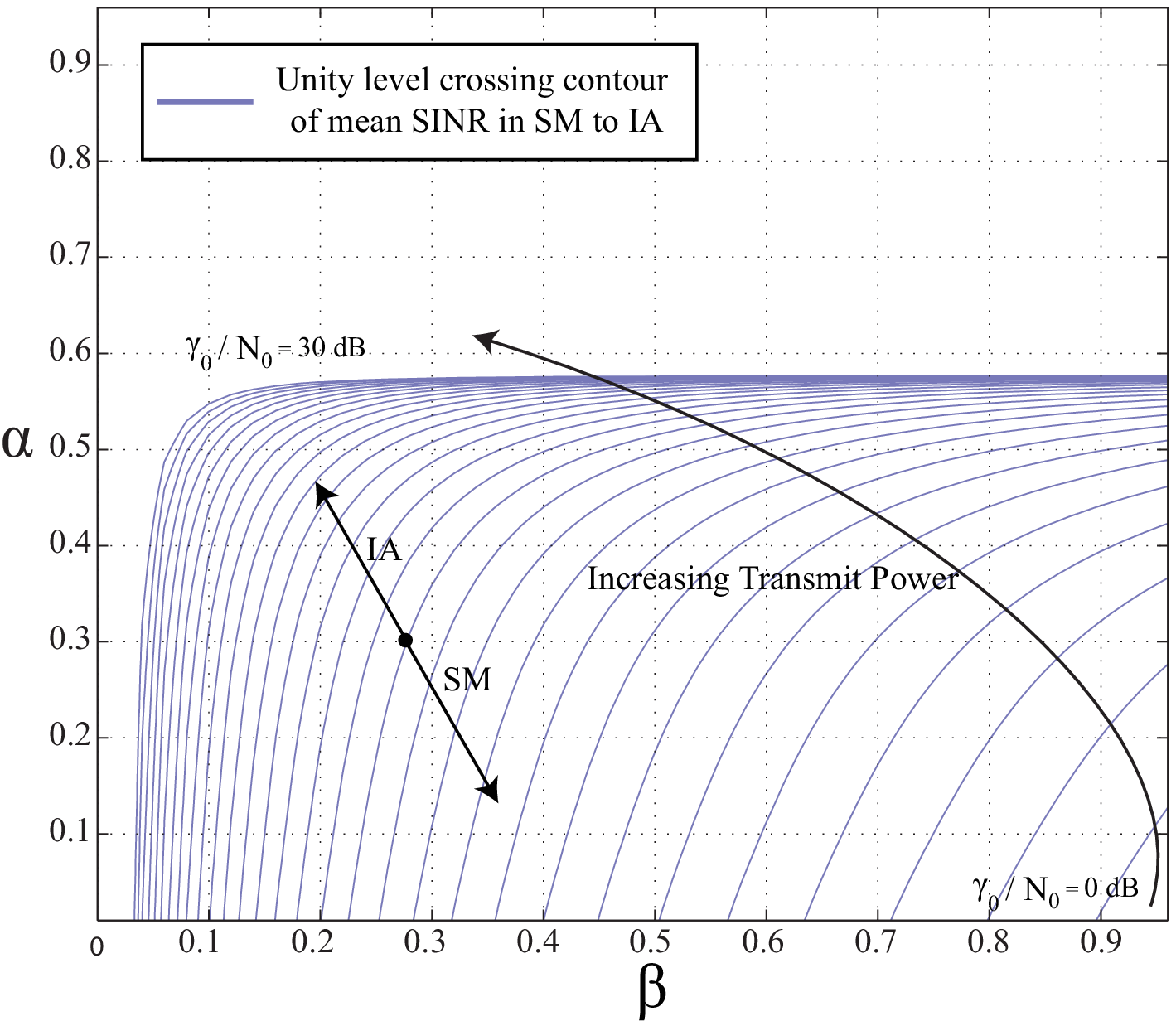}
\caption{Theoretical unity level contour plots for the ratio of the per stream mean SINR in a single-user $2\times 2$ SM MIMO link to per stream mean SINR of a $3$-user $2\times 2$ MIMO IA network given by (\ref{eq:RatioOfP2PtoIA}) for varying $\beta$, $\gamma_o$ and $\alpha$. As expected, the SM link has a better performance for large values of $\beta$ and small values of $\alpha$.}
\label{fig:BetaVsAlphaContourTheo}
\end{figure}

\begin{figure}[t]
 \centering
\includegraphics[bb= 0 0 410 360 ,width=4in]{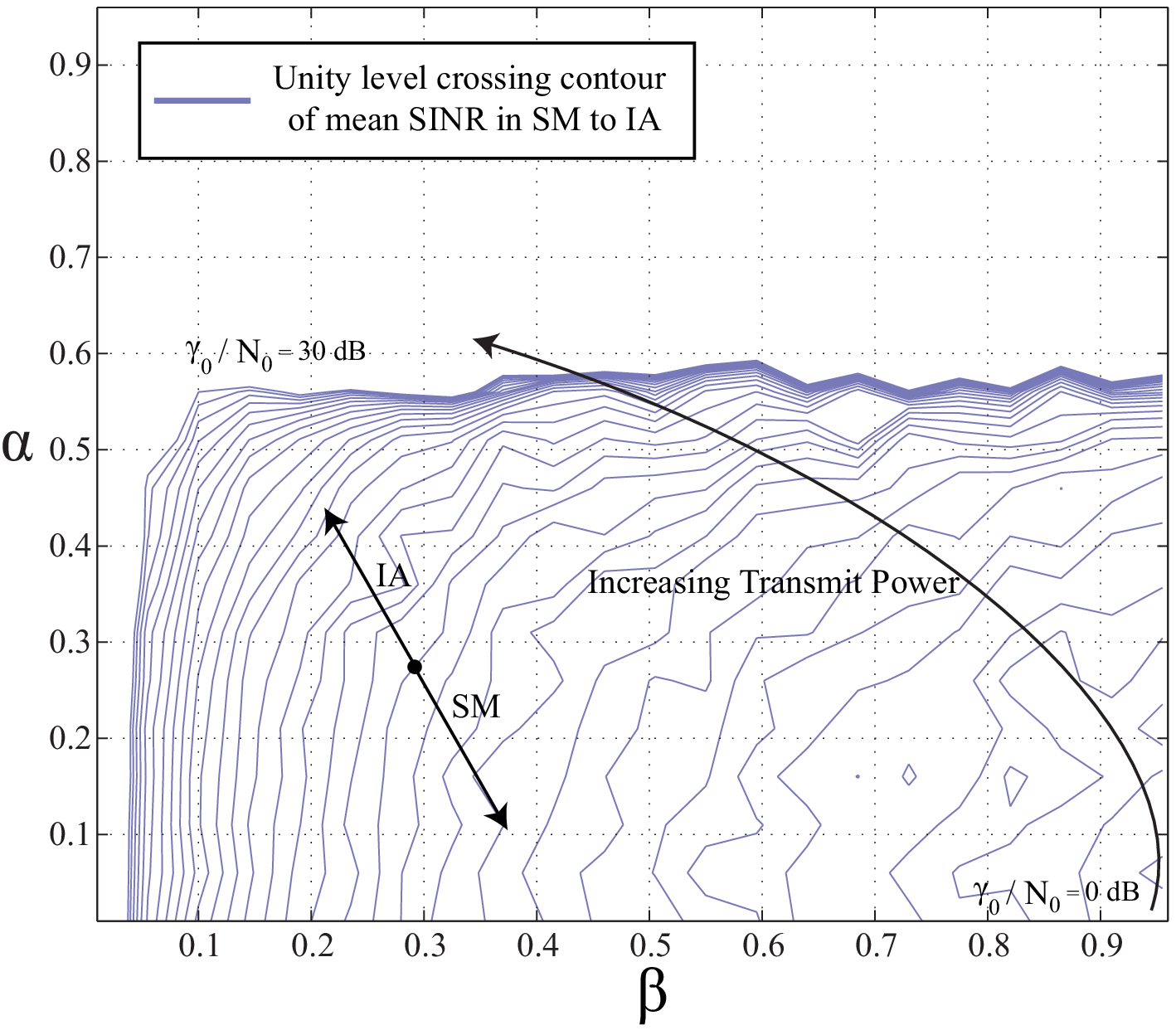}
\caption{Numerical unity level contour plots for the ratio of the per stream mean SINR in a single-user $2\times 2$ SM MIMO link to per stream mean SINR of a $3$-user $2\times 2$ MIMO IA network given by (\ref{eq:RatioOfP2PtoIA}) for varying $\beta$, $\gamma_o$ and $\alpha$. As expected, the SM link has a better performance for large values of $\beta$ and small values of $\alpha$.}
\label{fig:BetaVsAlphaContourSim}
\end{figure}

\indent Consider a $3$-user $2\times 2$ MIMO IA network with $d_i=1, i\in\{1,\ldots,K\}$ and a $2\times 2$ SM MIMO link with a ZF receiver. Although it can be shown that the $R_{\rm{sum}}$ of the SM link, in this configuration, is always less than the $R_{\rm{sum}}$ of the IA network, a measure of relative performance between the two networks can be defined as (\ref{eq:RatioOfP2PtoIA}). The contours curves of (\ref{eq:RatioOfP2PtoIA}), in terms of $\alpha$ and $\beta$ for varying $\gamma_o$, are shown in Fig. \ref{fig:BetaVsAlphaContourTheo}. As can be seen, the SM link has a better performance in the areas defined by large $\beta$ and small $\alpha$ and this area grows by increasing $\gamma_o$. This behavior can be explained by noting that the effect of imperfect CSI on the IA and SM networks is proportional to $K$ and $N$, respectively, and when $K>N$, imperfect CSI is more destructive for the IA network. The effect of antenna correlation, however, is more tolerated in the IA network where the eigenvalues of $\mathbf{R}_t$, unlike the SM MIMO link, are not directly limiting the ZF performance. The corresponding numerical curves of Fig. \ref{fig:BetaVsAlphaContourTheo} are shown in Fig. \ref{fig:BetaVsAlphaContourSim} where by increasing $\gamma_o$, the theoretical approximation better estimates the true behavior of the IA network. Moreover, both the theoretical and numerical contours show similar values for $\alpha$ and $\beta$ for which by increasing $\gamma_o$, the relative behavior of the SM link to IA network will not change, i.e. $\alpha \approxeq 0.58$ and $\beta \approxeq 0.04$, showing how our derived theoretical results can be used to accurately predict the IA system performance over a large range of $\beta$, $\alpha$ and $\gamma_o$.


\section{Conclusion}\label{sec:Conclusion}
\indent The performance of MIMO IA networks depend on the accuracy of CSI and channel correlation. This paper quantified the impact of imperfect CSI and transmit antenna correlation via the per-stream post-processing SINR distribution. Upon using zero-forcing equalizers in a Rayleigh channel, post-processing SINR was shown to be exponentially distributed with the mean value being a function of the number of antennas at each node, the transmit antenna correlation, the imperfection in CSI, and the transmit power. It was shown that, in the presence of imperfect CSI, the performance of IA degrades with increasing total number of streams in the network and if the imperfection does not vanish at asymptotically high transmit powers, the multiplexing gain is zero. Moreover, it was shown that as long as the channel matrices are full-rank, the impact of transmit correlation is less detrimental confined to a constant power loss which does not decrease the multiplexing gain achievable through IA. The performance of the two most commonly used transmit techniques in orthogonal access networks, beamforming and spatial multiplexing, was compared to the performance of IA by utilizing the derived SINR distributions where it was shown that IA is not always the optimum transmission strategy given realistic system parameters.   

\indent The results of this paper can be used as a starting point for further research. The performance of other communication techniques for interference channels such as combinations of orthogonal access, beamforming, coordinated multipoint transmission/reception and so on can be compared to the performance of MIMO IA networks and optimal switching points between different methods based on number of nodes, number of antennas, transmission power, amount of CSI imperfection or structure of the transmission correlation can be found which might help guide which techniques are appropriate for different networks.

\clearpage
\bibliographystyle{IEEEtran}
\bibliography{../../Bib/IEEEabrv,../../Bib/IA_main_ver2}

\newpage

\clearpage

\begin{table}[b]\label{table:AlphaValues}
\begin{center}
\caption{Channel correlation values for varying antenna spacing in a Sub-urban Macro-cell environment \cite{R4_060101}}
\begin{tabular}{|c||c|c|} 
 \hline  
Antenna Spacing ($\times \lambda$) & $\alpha$ & $|\alpha|$ 
\\ \hline \hline
10  & $-0.1743 + 0.0951i$ &  $0.1986$\\ \hline
9   & $0.2064 + 0.1066i$ &  $0.2323$\\ \hline
8   & $-0.0341 - 0.2872i$&  $0.2892$ \\ \hline
7   & $-0.2817 + 0.2408i$&  $0.3706$ \\ \hline
6   & $0.4551 + 0.1317i$&  $0.4738$ \\ \hline
5   & $-0.1717 - 0.5660i$&  $0.5915$ \\ \hline
4   & $-0.4616 + 0.5439i$&  $0.7134$ \\ \hline
3   & $0.8193 + 0.1101i$ &  $0.8267$\\ \hline
\end{tabular}
\end{center}
\end{table}

\end{document}